\documentclass[aps,pre,twocolumn,showpacs,superscriptaddress,groupedaddress]{revtex4-2}

\usepackage{amsmath}
\usepackage{amsfonts}
\usepackage{amssymb}
\usepackage{graphicx}
\usepackage{graphics}
\usepackage{dcolumn}
\usepackage{sidecap}
\usepackage{parskip}
\usepackage{xcolor}
\usepackage{hyperref}
\usepackage{hhline}
\usepackage{mathtools}
\usepackage{multirow}
\usepackage{verbatim}
\usepackage{rotating}
\usepackage{setspace}
\usepackage{epsfig}
\usepackage{epstopdf}
\usepackage{natbib}
\usepackage{subfigure}
\usepackage{booktabs}
\usepackage[normalem]{ulem}
\usepackage{bm}
\usepackage{verbatim}
\usepackage{comment}
\usepackage{cleveref}
\usepackage[normalem]{ulem}

\makeatletter
\def\@eqnnum{{\normalsize \normalcolor (\theequation)}}
 \makeatother

\graphicspath{{./}{ER/main/}}
\begin{document}
\title{Rotating clusters in phase-lagged Kuramoto oscillators with higher-order interactions}
%higher-order interactions}
\author{Bhuwan Moyal}
\altaffiliation{Both authors contributed equally.}
\author{Priyanka Rajwani}
\altaffiliation{Both authors contributed equally.}
\author{Subhasanket Dutta}
\author{Sarika Jalan}\email{sarika@iiti.ac.in}
\affiliation{Complex Systems Lab, Department of Physics, Indian Institute of Technology Indore, Khandwa Road, Simrol, Indore-453552, India}

\date{\today}

\begin{abstract}
The effect of phase-lag in pairwise interactions has been a topic of great interest for long. 
%An inclusion of phase-lag in coupld Kuramoto oscillators has been shown to delay the occurance of synchronization tranistion. 
However, real-world systems often have interactions that are beyond pairwise and can be modeled using simplicial complexes. 
%so to fully understand real-world systems, 
We investigate the impact of phase-lag in coupled Kuramoto oscillators with higher-order interactions and find that the inclusion of higher-order interactions shifts the critical point at which first-order transition from a cluster synchronized state to an incoherent state takes place.
%a generalized version of 2-simplex model with a phase lag. 
In the thermodynamic limit, by employing Ott-Antonsen approach we derive a reduced equation for the order parameter measuring cluster synchronization, and
%to achieve a closed expression for the order-parameter describing the synchronization profile, we 
%this problem 
%for the symmetric part using Ott-Antonsen ansatz, further to study the steady-state solutions 
progress further through the self-consistency method to obtain a closed form of the order parameter measuring global synchronization 
%of the global order parameter for triadic interactions, 
which was lacking in the Ott-Antonsen approach. %This reduced form describe the effect of triadic coupling strength (having phase lag) on synchronization. 
%Apart from setting different initial conditions, 
%The interaction strength for backward transition moves towards the right due to an increase in phase lag, 
Moreover, considering the polar coordinates framework we
obtain the rotation frequency of the clusters which 
comes out to be a function of the phase-lag parameter 
further indicating that phase-lag can be used as a control parameter to achieve a desired cluster frequency.
 %The work may have potential applications in understanding the origin of cluster synchronization prevalent in a range of real-world complex systems such as Brain, power grids,  and social contacts having higher-order interactions and phase-lag.

%In the following model, we have two synchronized cluster states where the inclusion of the phase lag parameter causes the transition from synchronization and asymmetry in two clusters to a completely incoherent state to occur at a high value of coupling strength causing frustration in the system.
\end{abstract}

\maketitle
\paragraph{\bf{Introduction:}}
Synchronization of interacting units occurs in many real-world complex systems ranging from the circadian clock in the brain, neural networks, power grids, cardiac rhythms, and chemical oscillators \cite{pikovsky}. It was the insights of Winfree that Kuramoto later utilized to model the collective phenomenon of synchronization into a more manageable form which tells us how the coupled Kuramoto oscillators progress from an incoherent to a fully coherent state through a second-order phase transition 
 \cite{kuramoto1975international, strogatz2000kuramoto}. Later studies on coupled Kuramoto oscillators model with and without a phase-lag have uncovered versatile phenomena, such as global and cluster synchronization \cite{rodrigues2016kuramoto}, partially synchronized state \cite{omelchenko2013bifurcations}, explosive synchronization in single layer  \cite{pazo2005thermodynamic} and multilayer networks \cite{ zhang2015explosive, khanra2018explosive, jalan2019inhibition} etc. Particularly, phase-frustrated coupling, which could also be perceived as time-delayed interactions is widespread in various physical systems \cite{wolfrum2022multiple, crook1997role, hsia2020synchronization}. For example, in power grids, the phase-lag parameter corresponds to the energy loss along the transmission lines 
 \cite{dorfler2012synchronization}. Also, a neural network with distributed time delays can be modeled as a coupled Kuramoto model with a phase-lag parameter \cite{duan2008development}.  Sakaguchi and Kuramoto investigated the effect caused by the inclusion of a phase frustration parameter in an ensemble of oscillators and uncovered that strong coupling of oscillators
 %coupled strongly enough that 
 bring them together in one cluster which rotates with a  non-zero frequency deviating from the algebraic sum of the oscillator's intrinsic frequencies \cite{sakaguchi1986soluble}. This contrasts with what was realized for the Kuramoto model having zero phase-lag. Further, phase-lag is found to be responsible for causing phase turbulence in self-oscillatory diffusive systems \cite{kuramoto1984cooperative}. %Moreover, apart from the usual behavior of the transition from an incoherent to a synchronized state through a partially synchronized state, the Sakaguchi-Kuramoto model displays non-trivial synchronization transitions where an increase in the coupling strength leads to a decrease in synchronization, in which incoherence starts regaining its stability, and partially synchronized, incoherent states coexist together \cite{omel2012nonuniversal}. 
 Few systems that have been modeled through Kuramoto oscillators with phase-lag parameters are seismology \cite{vasudevan2015earthquake}, Josephson junctions \cite{wiesenfeld1996synchronization, filatrella2000high}. Also, this model became the prototype model to investigate chimera where the non-locally coupled oscillators voluntarily split into synchronized and incoherent populations \cite{abrams2004chimera}.  

However, all these investigations and results were largely confined to purely pairwise interactions. Recent advances have indicated that this simplistic view might not be sufficient to fully decipher the underlying mechanisms behind many real-world complex phenomena where higher order or $n-$simplicial interaction exist \cite{ battiston2020networks, boccaletti2023structure}.  In complex systems, $n$-simplex is formed by $n+1$ interacting nodes, for example, $2-$ and $3-$ simplex represent triangle and tetrahedron, respectively \cite{baccini2022weighted}. Many real-world complex systems such as the Brain, scientific collaborations, and social systems have underlying higher-order interactions that are crucial for their functioning and evolution \cite{iacopini2019simplicial}. 
 In 2019, Skardal and Arenas \cite{skardal2019abrupt} had shown that 
 the Kuramoto oscillators coupled through $2-$simplex interactions manifest an abrupt first-order transition to de-synchronization with no complementary abrupt synchronization transition. Also, in the thermodynamic limit, there exist continuum de-synchronization transition points arising due to changes in the initial conditions. Later Kachhvah and Jalan \cite{kachhvah2022first} showed that adaptive $1-$ and $2-$ simplicial interactions can lead to first-order transition to anti-phase clusters. The past few years have witnessed remarkable growth in the studies of coupled Kuramoto oscillators with higher-order interactions on single layer network \cite{xu2021spectrum, gao2023dynamics, rajwani2023tiered, sabhahit2023self, anwar2023neuronal, adhikari2023synchronization}, furthermore on multilayer networks \cite{jalan2022multiple, rathore2023synchronization}. Lately, Carletti et al. \cite{carletti2023global} have shown that only some particular topological oscillators in higher-order networks exhibit global synchronization which is not seen in any arbitrary simplicial complexes.
 %{\color{red} also in \cite{adhikari2023synchronization} higher-order interactions with pairwise coupling has been abstracted to complex hypergraphs with separate links and triangle distributions}.

 \begin{figure*}[t!]
\begingroup
\begin{tabular}{c c}
     \includegraphics[width=0.28\textwidth]{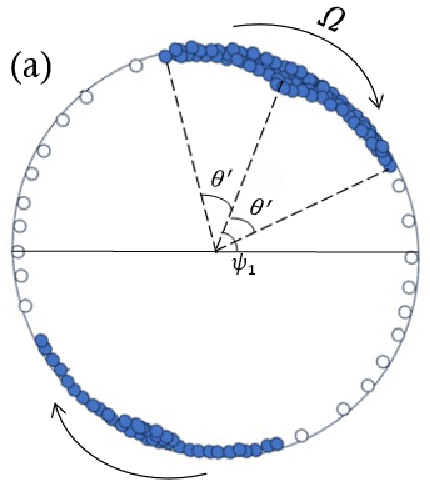}\hspace{1cm}
     \includegraphics[width=0.7\textwidth]{F1_b_c_eta_09_phi.eps}
        
     %{\includegraphics[width=0.37\textwidth]{alpha_eta.eps}}
\end{tabular}
\endgroup
\caption{(Color online) (a) Schematic diagram depicting fixed points of locked (solid circle) oscillators participating into rotating clusters along with drifting (open circle) oscillators. %{\color{red}\sout {Here, $\psi$ represents the average phase of the locked oscillators.}} 
(b) $r_1$ and, (c) $r_2$ vs $K_2$ for different $\alpha$ values depicting shift in critical coupling strength. Here, $\alpha=0$ (violet, open circles), $\pi/6$ (red, squares), $\pi/4$ (orange, right triangles), and $\pi/3.5$ (blue, left triangles). These results are obtained numerically by simulating Eq.~\ref{fin_model_Eq} adiabatically in the backward direction for $N=10^4$ oscillators. Solid curve lines represent analytical prediction from Eq.~\ref{fin_locked} and ~\ref{r2} using $\Omega$ values from Eq.~\ref{omega}. Solid circles correspond to numerical simulation in the forward direction.}
    \label{figl_alpha}
\end{figure*}

%Here we consider a $2$-simplex Kuramoto model with a phase lag parameter to analyze how the dynamics of synchronization get affected. 
A recent study on coupled Kuramoto oscillators model incorporating phase-lag parameter $\alpha$ in the triadic interactions along with the pairwise interactions has considered the following form of the triadic coupling $\sin(2\theta_j-\theta_k-\theta_i-\alpha)$, where $\theta_i$ is the phase of $i^{th}$ Kuramoto oscillator \cite{PhysRevE.108.034208}. Here, in this article, we consider another phase reduction form of
the complex Ginzberg Landau equation for 2-simplex interactions \cite{leon2019phase} yielding the triadic coupling as  $\sin(\theta_j+\theta_k-2\theta_i-\alpha)$. Such a form of the 2-simplex coupling in absence of a phase-lag is known to manifest two cluster state 
%, which gives rise to a continuum of 
that gets destroyed through an abrupt de-synchronization transition as coupling strength is adiabatically decreased 
%points as well as an infinite number of multistable states based upon different initial conditions 
 \cite{skardal2019abrupt}. Here, we show that the inclusion of a phase frustration term shifts the critical de-synchronization point toward the higher positive coupling strength value. That is, starting with a cluster-synchronized state, as coupling strength decreases adiabatically, de-synchronization to an incoherent state occurs for larger coupling strengths than that achieved for the zero phase-lag cases. The crucial difference between the form of the triadic interactions considered here from the form of the triadic interactions considered in reference {\color{red} \cite{PhysRevE.108.034208}} and other existing models having phase-lagged in the pairwise interactions {\color{red}\cite{sakaguchi1986soluble, omel2012nonuniversal}} is the existence of stable two-cluster state, in contrast to a stable global synchronized state. 
%{\color{red}Here two-cluster synchronization simply means that synchronization of locked oscillators is seen in $2$ different groups rather than one single cluster.} 
%and provides a more intrinsic view of the system.
%We only need two order parameters here to describe the synchronization state of the locked oscillators fully, as we will see the above model only allows for two clusters in the system}. 

In the thermodynamic limit, using the Ott-Antonsen approach \cite{ott2008low} we first derive the reduced dimensional equation for cluster synchronization state, and then by using the self-consistency method obtain the closed forms of order parameter corresponding to the global synchronization. The challenge lies in deriving an analytical expression of the cluster frequency ($\Omega$)  which comes out to be different from the mean of the intrinsic frequency of the oscillators and rather manifests an explicit dependence on $\alpha$, in contrast to the zero phase lag case.
%There exists an unique value of the mean frequency ($\Omega$) which 
%satisfies
%the self-consistency relation. , 
Ergo,  $\alpha$ can be used as a control parameter to regulate the rotation frequency of clusters to a desired value \cite{lohe2015synchronization}.
%similarly describe for phase frustrated Kuramoto model ($\sin(\theta_j-\theta_i-\alpha)$) \cite{lohe2015synchronization}. 
Further, we present the numerical simulations for finite-size networks which show a good match with the analytical predictions performed in the thermodynamic limit.

\paragraph{\bf{Model:}} 

We consider a higher-order extension of the Kuramoto-Sakaguchi model with coupling taken as 2-simplex interactions,
\begin{equation}\label{mode_eq}
\dot\theta_i=\omega_i+\frac{K_2}{N^{2}}\sum_{j=1}^{N}\sum_{k=1}^{N}\sin(\theta_j+\theta_k-2\theta_i-\alpha),
\end{equation}
where $\omega_i$ is the intrinsic frequency of $i^{th}$ oscillator and $K_2$ is the $2$-simplex coupling strength for $N$ oscillators. The collective behavior of the oscillators can be analyzed using the definition of the generalized order parameter $z_q={r_q}e^{\iota\psi_q}=\frac{1}{N}\sum_{j=1}^{N}e^{q\iota\theta_j}$ for $q=1,2$, where $r_1$ and $r_2$ measure the magnitude of global and two cluster synchronization, respectively. Two cluster synchronization refers to the state in which oscillators get locked in two cluster rather than one single cluster. Further, here $r_1$=$r_2$=0 indicates that oscillators are uniformly distributed in a circle of unit radius referred as incoherent state. Whereas $r_1=r_2=1$ implies the global synchronization in which all oscillators are locked in a single cluster. Another case of  $r_1=0$ and $r_2=1$ indicates anti-phase two cluster synchronization. The mean phase $\psi_q$ can be calculated as
\begin{equation}\label{psi_q}
    \psi_q=\arctan\left(\frac{\sum_{j=1}^{N}\sin(q\theta_j)}{\sum_{j=1}^{N}\cos(q\theta_j)}\right).
\end{equation}

\paragraph{\bf {Mean-field equation and analytical calculations:}} Order-parameter notions help us to write Eq.~\ref{mode_eq} in the mean-field form such as 
\begin{equation}\label{mean_field}
    \dot\theta_i=\omega_i+{K_2}{r_1^{2}}\sin(2\psi_1-2\theta_i-\alpha). 
\end{equation}
In the continuum limit $N \rightarrow \infty$ the state of the system can be given by density function $\rho(\theta,\omega,t)$ which describes the density of oscillators with phase between $\theta$ and $\theta+\delta\theta$ and intrinsic frequencies between $\omega$ and $\omega+\delta\omega$ at time $t$. Since the number of oscillators are conserved, $\rho$ must satisfy the continuity equation 
\begin{equation}\label{cont_eq}
 \frac{\partial \rho}{\partial t}=-\frac{\partial(\rho\dot\theta)}{\partial\theta}.
\end{equation}
Considering the frequency of each oscillator drawn from a distribution $g(\omega)$, the density function can be expanded into Fourier series
\begin{equation}\nonumber
\rho(\theta,\omega,t)=\frac{g(\omega)}{2\pi}\left(\sum_{n=-\infty}^{\infty}\rho_n(\omega,t)e^{{\iota}n\theta}\right), 
\end{equation}
where $\rho_n(\omega,t)$ being the $n^{th}$ Fourier coefficient $\rho_{-n}=\rho_{n}^*$ and $\rho_0(\omega,t)=1$. We can write the density function into the sum of the symmetric and anti-symmetric parts; $\rho_s(\theta+\pi,\omega,t)=\rho_s(\theta,\omega,t)$ and $\rho_a(\theta+\pi,\omega,t)=-\rho_a(\theta,\omega,t)$. The linearity property of the continuity equation suggests that individually $\rho_s$ and $\rho_a$ are solutions, therefore the linear combination of both is also a solution. However, only the symmetric part allows for dimensionality reduction using Ott-Antonsen ansatz as all the Fourier modes decay geometrically \cite{ott2008low}, i.e., $\rho_{2n}(\omega,t)=\upsilon^n(\omega,t)$ where $|\upsilon(\omega,t)|\le 1$, 
\begin{equation} \label{fou_exp}
\rho_s(\theta,\omega,t)=\frac{g(\omega)}{2\pi}\left[1+\sum_{n=1}^{\infty}\rho_{2n}(\omega,t)e^{in\theta}+c.c\right].
\end{equation}
Plugging this and Eq.~\ref{mean_field} into the continuity Eq.~\ref{cont_eq}, we find that each subspace spanned by odd terms $e^{\iota n \theta}$ does not collapse into a low-dimensional manifold.  Whereas, the subspace defined by even term $e^{2\iota n \theta}$ does, i.e., given as,
\begin{equation}\label{mid_ott}
    \frac{\partial\upsilon}{\partial t}= -2\iota\upsilon\omega+K_2({z^{*}_1}^{2}{e^{\iota\alpha}}-z_1^{2}\upsilon^{2}e^{-\iota\alpha}).
\end{equation}
In the continuum limit $N\rightarrow\infty$, we have $z_2=\int_{-\infty}^{\infty}\int_{0}^{2\pi}{\rho_s(\theta,\omega,t)e^{2\iota \theta}d\theta{d\omega}}$, which after inserting the Fourier series expansion of $\rho_s(\theta,\omega,t)$ reduces to $z_2=\int_{-\infty}^{\infty}g(\omega){\upsilon^*}d\omega$. Upon considering the frequency distribution $g(\omega)$ to be Lorentzian   {\large$g(\omega)=\frac{\Delta}{\pi[(\omega-\omega_0)^2+\Delta^2]}$} with mean $\omega_0=0$ and spread $\Delta=1$, the complex integral ${z_2}$ can be calculated using Cauchy's residue theorem by contour integration in the negative half-plane, yielding $z_2={\upsilon^*}(\omega_0-\iota\Delta,t)$. Further, taking complex conjugate of Eq.~\ref{mid_ott} and substituting $\omega=\omega_{0}-\iota\Delta$,
%After making this substitution, {\color{red} and using the definition of $z_2$, $z_1$},
\begin{equation}\nonumber
  \frac{\partial z_{2}}{\partial t}=-2z_{2}+K_{2}({z_{1}}^2 e^{-\iota\alpha}-{z^{*}_{1}}^2 z^{2}_{2} e^{\iota\alpha}).  
\end{equation}
Upon employing the definition of $z_2$ and $z_1$, while separating the real and imaginary parts, Eq.~\ref{mid_ott} reduces to
\begin{equation}\label{ana1}
    \dot{r_2} = -2r_2 +K_2{{r_1}^{2}}(1-r_2^{2})\cos(2\psi_1-\psi_2-\alpha).
\end{equation}
\begin{equation}\label{ana2}
    \dot\psi_2 = K_2{r_1^{2}}\frac{1+r_2^{2}}{r_2}\sin(2\psi_1-\psi_2-\alpha).
\end{equation}
Note that these equations are achieved by considering the contribution of the symmetric part  ($\rho_s$) only, which does not accomplish an explicit relation between $r_1$ and $K_2$, %{\color{red}since the subspace spanned by odd terms $e^{\iota n \theta}$ dose not collapsed into the low-dimensional manifold.}
Hence, we proceed further with the self-consistency method.

%Proceeding further to analyze the $z_1$ using the self-consistency method. 
We change the frame of reference $\theta \rightarrow \theta + \psi_1$, and enter into the rotating frame of the cluster ($\dot\psi_1=\Omega$), %{\color{red} \sout {which yields $\psi_1$ and $\psi_2$  zero \cite{xu2020bifurcation}.} }
Hence, Eq.~\ref{mean_field} can be written as 
\begin{equation}\label{rot_mean_field}
    \dot\theta_i=\omega_i-\Omega-K_2{r_1}^2\sin(2\theta_i+\alpha).
\end{equation}
Note that when $\alpha=0$, oscillators are distributed in a complex circle around the mean $\psi_1$ guided by the frequency distribution $g(\omega)$. Also, on changing the value of $K_2$ the frequency range of the locked oscillators participating in clusters remains symmetric about zero. However, for non-zero $\alpha$ values effective clusters frequency for $K_2>K_{2c}$ (i.e., the critical coupling strength where the transition occurs) will be different from the mean of intrinsic frequencies.  Consequently, the synchronized clusters rotate with a common non-zero frequency $\Omega$ with the magnitude of the maximum frequency being different from that of the $\alpha=0$ case. 
\begin{figure}
    \centering
    \begingroup
  \renewcommand{\arraystretch}{1.5}
   % \begin{tabular}{c}
     %   \centering
        \includegraphics[width=\linewidth]{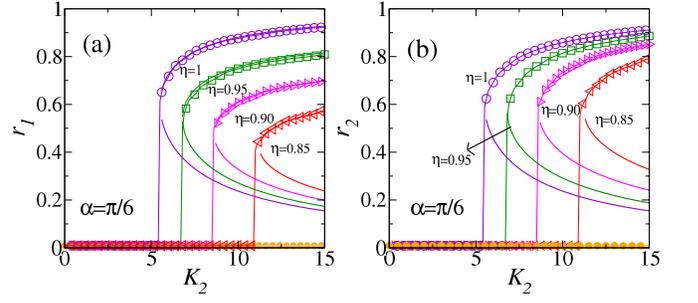}
    
         %\includegraphics[width=\linewidth]{eta_09_phi.eps}
        
    %\end{tabular}
    \endgroup
       
        % \caption{Caption for the second subfigure} % You can add captions if needed
         
   \caption{(Color online) Synchronization profiles depicting, $r_1$, and $r_2$ as a function of $K_2$: (a) and (b) are for different $\eta$ values. Here, $\eta=1$ (violet, open circles), $0.95$ (green, squares), $0.90$ (magenta, right triangles), and $0.85$ (red, left triangles). These results are obtained numerically by simulating Eq.~\ref{fin_model_Eq} adiabatically in the backward direction. Solid lines correspond to analytical prediction obtained by Eq.~\ref{fin_locked} and ~\ref{r2} using $\Omega$ values from Eq.~\ref{omega}}. Solid circles correspond to numerical simulation in the forward direction. 
   \label{fig2_eta}
\end{figure}

Next, the whole population can be divided into two groups of the locked and drifting oscillators such as $|{\frac{\omega_i - \Omega}{K_2r_1^2}}| \le 1$ and $|{\frac{\omega_i - \Omega}{K_2r_1^2}}|>1$, respectively. Moreover, for the locked oscillators, the coupling form of the higher-order interactions considered in Eq.~\ref{rot_mean_field} renders two stable fixed points;
\begin{equation}\label{fix_point}
\theta^*=\frac{1}{2}\arcsin\left(\frac{\omega_i - \Omega}{K_2r_1^2}\right)-\frac{\alpha}{2},\,\,\,\,\,\,\,\,\,\,   \theta^*+\pi. 
\end{equation}
 This indicates the presence of two cluster synchronization. Since $r_1$ cancels out the presence of the number of oscillators in both clusters and measures effective synchronization. To quantify two cluster synchronization, we have defined $z_2={r_2}e^{\iota\psi_2}=\frac{1}{N}\sum_{j=1}^{N}e^{2\iota\theta_j}$; where the value of $r_2$ will measure the extent of two cluster synchronization.
Further, to study the contribution of the locked oscillator the density function can be written as, 
\begin{equation}
    \rho_{lock}(\theta,\omega)=\eta \delta(\theta-\theta^*)+(1-\eta)\delta(\theta-(\theta^*+\pi)),
\end{equation}
where $\eta$ and $1-\eta$ depict the probability of oscillators and  having value $\theta^*$ and $\theta^*+\pi$, respectively. Here, $0<\eta<1$ signifies the fraction of locked oscillators present in the $\theta^{*}$ cluster. Furthermore, $z_1=\int_{-\infty}^{\infty}\int_{0}^{2\pi}{e^{\iota \theta}\rho_{loc}(\theta,\omega)g(\omega)d\theta{d\omega}}$ provides the contribution from the locked oscillators given by,
\begin{equation}\label{locked}
    {r_1}^{lock}=(2\eta-1)\int_{-K_{2}r_1^{2}+\Omega}^{K_{2}r_1^{2}+\Omega} e^{\iota\theta^*}g(\omega)d\omega.
\end{equation}
Moreover, for the locked state $\dot \theta=0$ (Eq.~\ref{rot_mean_field}), 
Using the trigonometric identities, the above equation can be expressed as 
\begin{equation}\nonumber
\begin{split}
    \cos(\theta^*+\frac{\alpha}{2})&=\sqrt{\frac{1+\sqrt{1-(\frac{\omega_i - \Omega}{K_2r_1^2})^2}}{2}},\\
     \sin(\theta^*+\frac{\alpha}{2})&=\pm \sqrt{\frac{1-\sqrt{1-(\frac{\omega_i - \Omega}{K_2r_1^2})^2}}{2}},
\end{split}
\end{equation}
where $\theta^*=\Theta-\frac{\alpha}{2}$. The contribution of the sinusoidal term will be either positive or negative based on the limits of the integration over $\omega$. Hence, Eq.~\ref{locked} can be expressed as 
\begin{equation}\nonumber
    {r_1}^{lock} e^{\iota \frac{\alpha}{2}}=(2\eta-1)\int_{-{K_2{r_1}^2}+\Omega}^{{K_2{r_1}^2}+\Omega} e^{\iota \Theta} g(\omega)d\omega.
\end{equation}
By plugging the value of $\Theta$ and comparing the real and imaginary parts, the  contribution from the locked oscillators gets determined as 
\small{
\begin{align}\label{fin_locked}
{r_1}^{lock}=&(2\eta-1)[\cos{\frac{\alpha}{2}}\int_{-K_2{r_1}^2+\Omega}^{K_2{r_1}^2+\Omega}\sqrt{\frac{1+\sqrt{1-{(\frac{\omega-\Omega}{K_2{r_1}^2})}^2}}{2}}g(\omega)d\omega \nonumber \\
    &- \sin{\frac{\alpha}{2}}\int_{-K_2{r_1}^2+\Omega}^{\Omega}\sqrt{\frac{1-\sqrt{1-{(\frac{\omega-\Omega}{K_2{r_1}^2})}^2}}{2}}g(\omega)d\omega \nonumber \\
   & + \sin{\frac{\alpha}{2}}\int_{\Omega}^{K_2{r_1}^2+\Omega}\sqrt{\frac{1-\sqrt{1-{(\frac{\omega-\Omega}{K_2{r_1}^2})}^2}}{2}}g(\omega)d\omega ].
\end{align}
}
In addition, to analyze the contribution of the drifting oscillators, where ${|{\frac{\omega_i - \Omega}{K_2r_1^2}}|>1}$
\begin{equation}\label{drift}
    {r_1}^{drift}=\int_{|{\frac{\omega - \Omega}{K_2r_1^2}}|>1}^{} \int_{0}^{2\pi}e^{\iota\theta}\rho_{drift}(\theta,\omega)g(\omega)d\omega{\color{red}d\theta},
\end{equation}
from Eq.~\ref{cont_eq} in the steady state, $\rho_{drift}\dot\theta$ should be a constant yielding $\rho_{drift}=\frac{C}{\dot\theta}$, where the normalization constant calculated as the total probability of finding the oscillators on a circle with $\int_{-\pi}^{\pi}{\rho_{drift}(\theta,\omega)d(\theta)}=1$ for each $\omega$.
\begin{equation}
\label{driftdensity}
   \rho_{drift}=\frac{\sqrt{(\omega-\Omega)^2-({K_2{r_1}^2})^2}}{2\pi|\omega-\Omega-{K_2{r_1}^2}\sin(2\theta+\alpha)|}.
\end{equation}
It is clear from Eq.~\ref{driftdensity}, $\rho_{\color{red}{drift}}(\theta,\omega)=\rho_{\color{red}{drift}}(\theta+\pi,\omega)$, which yields $r_1^{drift}=0$ (Eq.~\ref{drift}), i.e., contribution of the drifting oscillators in $r_1$ vanishes. Therefore, $r_1=r_1^{lock}+r_1^{drift}\approx r_1^{lock}$. 
 Furthermore, from Eq.~\ref{ana1}, we get the fixed point solution of $r_2$ as,
\begin{equation}\label{r2}
    r_2=\frac{-1+\sqrt{1+{{K_2}^2{r_1}^4{\cos^2{\alpha}}}}}{{K_2{r_1}^2\cos{\alpha}}}.
\end{equation}

Further, to simplify the expression of the self-consistency Eq.~(\ref{fin_locked}) we have to determine the rotation frequency $\Omega$. 
%The general definition for $\psi_q$ can be given as,
%\begin{equation}
%\psi_q=\arctan(\frac{\sum_{j=1}^{N}\sin(q\theta_j)}{\sum_{j=1}^{N}\cos(q\theta_j)}).
%\end{equation}
\begin{figure*}
\begingroup
\begin{tabular}{c c}
    
    \includegraphics[width=0.60\textwidth]
    {F3_a_b_lorentzian.eps}
    \includegraphics[width=0.31\textwidth]{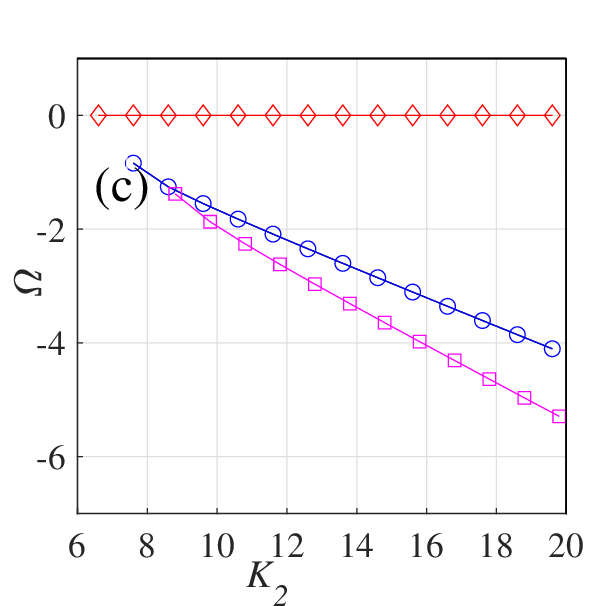}
    \end{tabular}
    \caption{Schematic diagram depicting the range of the oscillators taking part in the two cluster state for (a) $\alpha=0$ case, and (b) for $\alpha\neq0$ value illustrates that oscillators taking part in the synchronized clusters may not be symmetric about mean 0. (c) $K_2$ vs $\Omega$ for $\eta=0.9$ and different values of $\alpha$ such as $0$ (red, diamonds), $\pi/8$ (blue, circles), and $\pi/6$ (magenta, squares) plotted using Eq.~\ref{omega}.  } 
    \label{Omega vs K2}
    \endgroup
\end{figure*}
The summation in Eq.~\ref{psi_q} can be written in terms of the contribution from the locked and drifting oscillators,
$\psi_q=\arctan(\frac{\langle \sin(q\theta)\rangle_{lock} +\langle \sin(q\theta)\rangle_{drift}}{\langle \cos(q\theta)\rangle_{lock}+\langle \cos(q\theta)\rangle_{drift}})$. Oscillators lie in two cluster having $\pi$ phase difference (Eq.~\ref{fix_point}). It is straightforward now to see that contributions from the drifting oscillators cancel out due to symmetric density distribution (Eq.~\ref{driftdensity}), as depicted in Fig.~\ref{figl_alpha}(a).
To calculate the arguments of $\psi_{2}$, we first expand  $\langle \sin(q\theta)\rangle_{lock}$ term in one cluster which are $\eta[sin({2}\theta_{1})+sin({2}\theta_{2})+......+sin({2}\theta_{n-1})+sin({2}\theta_{n})]$ and in another cluster, $(1-\eta)[sin({2}(\theta_{1}+\pi))+sin({2}(\theta_{2}+\pi))+......+sin({2}(\theta_{n-1}+\pi))+sin({2}(\theta_{n}+\pi))]$ upon adding and solving we obtain,
\small{
\begin{align}\nonumber
\psi_{2}=\arctan\left(\frac{sin{2}\theta_{1}+sin{2}\theta_{2}+......+sin{2}\theta_{n-1}+sin{2}\theta_{n}}{cos{2}\theta_{1}+cos{2}\theta_{2}+......+cos{2}\theta_{n-1}+cos{2}\theta_{n}}\right).
    \end{align}
}

%Furthermore, we will define $\psi$ as the average of all the phases in a particular cluster indicated in Fig.~\ref{figl_alpha}(a)}. 
Then, for the locked oscillators in a cluster lying at, $\psi_{\color{red}1} + \theta^{'}$ there will exist another locked oscillator in the same cluster lying at $\psi_{\color{red}1}-\theta^{'}$ since the intrinsic frequency distribution of oscillators is symmetric (Fig.~\ref{figl_alpha}(a)).
 %Again, by rewriting the arguments,  the numerator becomes
% $$(sin(\psi-\theta)+sin(\psi+\theta)+sin(\psi-\theta^{'})+sin(\psi+\theta)+.....)$$, where $\theta$ is the angle between mean and boundary of cluster, and going towards the center the angle between next nearest neighbors and center is $\theta^{*}$. 
Further, simplifying the expression we get,
 \begin{equation}\nonumber
    \psi_{2}=\arctan\left(\frac{sin{2}{\psi_1}(2(cos{\theta}+cos\theta^{'}+...))}{cos{2}{\psi_1}(2(cos{\theta}+cos\theta^{'}+...))}\right). 
 \end{equation}
Thus, we obtain $\psi_2=2\psi_1$.
% \begin{equation}
%$\psi_q=\arctan(\frac{\sin(q(\psi+\theta^{'}))+\sin(q(\psi+\theta^{''}))+....+\sin(q(\psi-\theta{''}))+\sin(q(\psi-\theta^{'}))}{\cos(q(\psi+\theta^{'}))+\cos(q(\psi+\theta^{''}))+....+\cos(q(\psi-\theta^{''}))+\cos(q(\psi-\theta^{'}))})$, 
%where $\psi=\frac{\sum_{j=1}^{N}\theta_j}{N}$. \textbf{$\theta^{'}$,  $\theta^{''}$ are phase differences of the locked oscillators lying above (below) with the mean phase ($\psi$)}.
This enables us to provide an important relation between the mean phase of global and cluster synchronization, which is necessary to calculate $\Omega$ as Ott-Antonsen only allows to get relation for even part of the dynamics. Next, using Eq. ~\ref{ana2} and above relation, we get $\dot{\psi_1}=\Omega$ such as,
\begin{equation}\label{omega}
    \Omega=-\tan{\alpha}\sqrt{(1+{{K_2}^2{r_1}^4{\cos^2{\alpha}}})}.
\end{equation}
 Next by substituting $\Omega$ from Eq.~\ref{omega}, the analytical values of $r_1$ are found by numerically integrating Eq.~\ref{fin_locked} for fixed $\alpha$, $K_2$ and $\eta$. Note that only the locked oscillators contribute in $r_1$ ($r_1^{drift}=0$, Eq.~\ref{drift}). Thereafter, we calculate $r_2$ from Eq.~\ref{r2} by inserting  values of $r_1$.

\paragraph{\bf {Numerical Calculations:}}
Further, by using the Eq.~\ref{omega}, mean field Eq.~\ref{rot_mean_field} reduces to
\begin{equation} \label{fin_model_Eq}
    \dot\theta_i=\omega_i+ \tan\alpha\sqrt{(1+{{K_2}^2{r_1}^4{\cos^2{\alpha}}})}-K_2{r_1}^2\sin(2\theta_i+\alpha).
\end{equation}
This equation incorporating a change of the reference frame (free of $\psi_1$) enables us to get rid of the rotation of the cluster. We numerically simulate the Eq.~\ref{fin_model_Eq} in a rotating cluster frame free of $\psi_1$, instead of Eq.~\ref{mean_field} for $N=10^{4}$. RK-4 method is used with a time step $=0.05$, and $r_1$ and $r_2$ are obtained by averaging over $2*10^{4}$ iterations after removing the initial transient period. We would like to note that though results from the numerical simulations of Eq.~\ref{mode_eq} and Eq.~\ref{fin_model_Eq} are same Fig.~\ref{A1_figl_K2_eta_alpha}(a), generating simulation results for Eq.~\ref{mode_eq} is a time-consuming process.
 %Numerical results are plotted against the 
 
 %obtained from the self-consistency and Ott-Antonsen analysis. 

\paragraph{\bf{Results:}}
  Fig.~\ref{figl_alpha}(a) represents the schematic distribution of oscillators in a complex unit circle. The figure illustrates that the locked oscillators are symmetrically distributed on a unit circle following the nature of the Lorentz distribution of the intrinsic frequency. There exist two anti-phase clusters with the locked oscillators density being $\eta$ and $1-\eta$, respectively rotating with a common angular frequency $\Omega$. Fig.~\ref{figl_alpha} (b) and (c) delineates $r_1$ and $r_2$ as a function of $K_2$ for different $\alpha$ values and $\eta=0.9$. The value of $r_2$ remains greater than $r_1$  portraying two cluster synchronization. Analytical predictions are obtained for specific values of $K_2$, $\eta$, and $\alpha$. The solution for $r_1$ is derived from Eq.~\ref{fin_locked} by employing values of $\Omega$ from Eq.~\ref{omega}. Additionally, the solution for $r_2$ is determined from  Eq.~\ref{r2} after substituting the obtained values of $r_1$.  
   There exists no forward synchronization (as $r_1$, $r_2$ $\rightarrow 0^{+}$) which is also supported if we invert Eq.~\ref{r2} which yields the forward critical coupling at $\infty$, thus upon increasing $K_2$ system always remains incoherent. 
%However, the backward critical coupling can be calculated by finding the minima of the analytical curve $dK_2/dr_1$ which is not possible to be determined explicitly in a closed form due to the complexity of integral in Eq.~\ref{fin_locked}. 
It can be seen that with an increase in $\alpha$ the backward transition point  ($K_{2c}$)  for $r_1$ and $r_2$ both shift towards the right. That is, the transition to the incoherent state occurs at a higher critical coupling value. 
As it happens that a non-zero $\alpha$ value yields a non-zero mean frequency (Eq.~\ref{omega}), the intrinsic frequency range of the locked oscillators satisfying the relation ($|{\frac{\omega_i - \Omega}{K_2r_1^2}}| \le 1$) no more remains symmetric around mean 0.  %As a result, the relation between $\psi_1$ and $\psi_2$ comes out to be linear.
 
Fig.~\ref{fig2_eta} (a) and (b), respectively present results for $r_1$ and $r_2$ as a function of $K_2$ for different values of $\eta$ at a fixed $\alpha$ value. To analyze the nature of phase transition to synchronization, we adiabatically increase and decrease $K_2$ representing the forward and backward direction, respectively. In the forward direction, initially, all the oscillators are distributed uniformly between [-$\pi$,$\pi$] and frequencies are drawn from a Lorentzian distribution. 
  Whereas in the backward direction, initially the oscillators are distributed into two clusters situated at diametrically opposite ends described by $\eta$.  As expected, in the absence of any pairwise couplings, there exists no forward synchronization for any $K_2$ value, whereas the backward direction yields a first-order transition from the cluster synchronized state to the incoherent state. With a decrease in $\eta$ critical transition point from the synchronized to the incoherent state shifts towards the right. 
  % I hvae corrected until here
  Additionally, as $\eta$ decrease, the oscillators having initial phase lying in the locked state attempt to distribute them in diametrically opposite ends which effectively renders less number of oscillators to contribute in $r_1$, due to which decreasing $\eta$ leads to an increase in the transition points for both $r_1$ and $r_2$, with $r_2$ being greater than $r_1$ for $\eta < 1$, depicting cluster synchronization. In the thermodynamic limit, multistable branches exist as an infinite number of stable partially synchronized states are obtained through different arrangements of the initial conditions in two different clusters, yielding a continuum of abrupt de-synchronization transitions. Moreover, we have shown the parameter space plot ($\eta$ vs $K_{2c}$) for $\alpha=\pi/8$ \&  $\alpha=\pi/6$, Fig.~\ref{A1_figl_K2_eta_alpha}(b).
  %Fig.~\ref{fig2_eta} (c) and (d) present behaviour of $r_1$ and $r_2$, respectively, as a function of $K_2$ for different $\alpha$ values. 
%There exists no forward synchronization (as $r_1$, $r_2$ $\rightarrow 0^{+}$) which is also supported if we invert Eq.~\ref{r2} which yields the forward critical coupling at $\infty$, thus upon increasing $K_2$ system always remains incoherent. 
%However, the backward critical coupling can be calculated by finding the minima of the analytical curve $dK_2/dr_1$ which is not possible to be determined explicitly in a closed form due to the complexity of integral in Eq.~\ref{fin_locked}. 
%It can be seen that with an increase in $\alpha$ the backward transition point  ($K_{2c}$)  for $r_1$ and $r_2$ both shift towards the right. That is, the transition to the incoherent state occurs at a higher critical coupling value. 
%As it happens that a non-zero $\alpha$ value yields a non-zero mean frequency (Eq.~\ref{omega}), the intrinsic frequency range of the locked oscillators satisfying the relation ($|{\frac{\omega_i - \Omega}{K_2r_1^2}}| \le 1$) no more remains symmetric around mean 0. 

Fig.~\ref{Omega vs K2}(a), the Lorentzian distribution considered here, the symmetry-breaking around the mean arising due to the inclusion of $\alpha$ will lead to less number of oscillators (Fig.~\ref{Omega vs K2}(b)). {Consequently, less number of oscillators contribute to the locked state with an increase in $\alpha$.} Fig.~\ref{Omega vs K2} (c) plots the rotation frequency of the clusters as a function of $K_2$. For $\alpha=0$ the cluster remains stationary for $K_2$ values yielding $\Omega=0$. However, for non-zero alpha values, $\Omega$ manifests a linear dependence on $K_2$ with an increasing slope (Eq.~\ref{omega}) even for the mean intrinsic frequency being zero. This demonstrates that the phase-lag parameter regulates the rotation frequency of the synchronized clusters which can be adjusted to a desired value by changing $\alpha$. A similar phenomenon is demonstrated for pairwise interactions with phase-lag but for the global synchronization \cite{lohe2015synchronization}. The crucial difference of the model considered here having triadic interactions from the pairwise interactions is that the former case yields $ 2-$ clusters in contrast to global synchronization in the latter case. 

\paragraph{\bf {Conclusion and outlook:}}
To conclude, we have analyzed the effects of phase frustration parameters on the coupled Kuramoto oscillators on simplicial complexes. We evaluated $r_1$ and $r_2$ order parameters which measure the extent of global and 2-cluster synchronization, respectively. In the absence of any pairwise interactions, $r_1=r_2=0$ remains one stable state for all $K_2$ values. Starting with a set of initial conditions corresponding to a synchronized state, as $K_2$ decreases adiabatically, there exists an abrupt transition to a completely incoherent state. With an increase in the $\alpha$ value, this transition point for both $r_1$ and $r_2$ shifts towards the right. Further, using the Ott-Antonsen dimension reduction approach we derived the time-dependent order parameter equations for the even part of the density function. To obtain the closed form of the asymmetric part, we proceed by self-consistency method which provides a relation between the order parameters (measuring global and cluster synchronization) and $K_2$. Additionally, to obtain the solutions for $r_1$ and $r_2$ we require an explicit expression of cluster frequency $\Omega$, which is not achievable through Ott-Antonsen ansatz or by self-consistency relation Eq.~\ref{fin_locked}. We propose an analytical method for determining the expression of $\Omega$, revealing an explicit dependence on $\alpha$. Therefore, $\alpha$ can serve as a control parameter to adjust the rotation frequency of clusters to a desired value. The analytical results are noted to be in good agreement with the numerical results. Also, dependence of the mean cluster frequency on $\alpha$ provides explanation behind the origin of non-zero mean cluster frequency even for intrinsic frequency distribution having zero mean. 
%Ergo, $\alpha$ can be used as a parameter to get the desired mean frequency of the synchronized clusters. 
%Moreover, for a fixed non-zero $\alpha$ value, multistable states are shown to exist (Fig.~\ref{figl_alpha}(b) and (c))
%where information on the arrangement of clusters is preserved  similar to the $\alpha=0$ case \cite{Skardal_prl2019}. 

This model can be generalized by including phase-lagged pairwise term along with the $2-$simplex interaction for which the self-consistency analysis becomes more challenging. Moreover, this model can be extended to multilayer networks. As demonstrated by Jalan and Suman \cite{jalan2022multiple} that multilayer networks can exhibit multiple first-order transition points instead of a single transition point to global synchronization. It will be inquisitive to investigate if phase-lagged higher-order interactions will lead to multiple first-transition to cluster synchronization. Further, there have been recent attempts to analyze coupled Kuramoto oscillators with inertia on simplicial complexes  \cite{sabhahit2023self}. An extension of the current work is to develop an analytical framework for the coupled Kuramoto model with inertia having phase-lag \cite{PhysRevE.108.024215}, which makes the model more generalized and suitable for wider applications.
\\
\\
\paragraph*{\bf{Access to Code:}} The source code for this study is available upon request.

\section*{\bf Acknowledgement}
SJ gratefully acknowledges SERB Power grant SPF/2021/000136, and useful discussions with Stefano Boccaletti under the VAJRA project VJR/2019/000034. PR is thankful to Govt of India, PMRF Grant
[No. PMRF/2023/2103358] 

\section*{Appendix}
\label{Appendix A}
\setcounter{figure}{0}
\setcounter{equation}{0}
\renewcommand{\thefigure}{A-\arabic{figure}}
\renewcommand{\theequation}{A-\arabic{equation}}
\begin{figure}[b!]
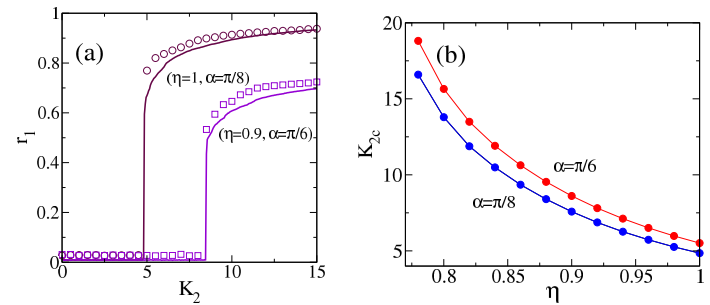

\begingroup
 \begin{tabular}{c c}
     \includegraphics[width=0.48\linewidth]{S1_a_sum_eta_alpha.eps}\hspace{0.25cm}
      \includegraphics[width=0.54\linewidth]{S1_b_k2_eta.eps}
    \end{tabular} 
\endgroup
\caption{(a) $K_2$ vs $r_1$ plot using direct simulation of Eq.~\ref{mode_eq} for $N=1000$ oscillators (symbol points) and using mean field Eq.~\ref{fin_model_Eq} by employing $\Omega$ values from Eq.~\ref{omega} for $N=10000$ oscillators (line). (b) Critical coupling transition points ($K_{2c}$) as a function of $\eta$ for different $\alpha$ values. Results obtained analytically through Eq.~\ref{fin_locked} by using $\Omega$ values from Eq.~\ref{omega}.}
    \label{A1_figl_K2_eta_alpha}
\end{figure}

\paragraph*{\bf{Critical coupling strength $(K_{2c})$ in parameter space:}} We observe that, with fixed value of $\alpha$, $K_{2c}$ is shifting towards higher positive values as $\eta$ decreases. Furthermore, keeping $\eta$ constant, an increase in $\alpha$ leads to $K_{2c}$ shifts towards higher positive values Fig.~\ref{A1_figl_K2_eta_alpha}(b).
% Your content for the introduction goes here

%\bibliographystyle{plainnat}
%\bibliographystyle{apsrev4-1}
%\bibliography{ref}

\end{document}